\documentclass[superscriptaddress,keywords]{revtex4}
\usepackage{epsfig,dcolumn}
\usepackage{graphicx}
\usepackage{color}
\usepackage{amsmath}
\usepackage{amsfonts}
\usepackage{amssymb}
\usepackage{graphicx}

\newcommand{\be}{\begin{equation}}
\newcommand{\ee}{\end{equation}}

\usepackage{bm} 

\begin{document}

\title{Dispersive calculation of 
complex Regge trajectories for the lightest $f_2$ resonances}  

\author{ J.A.~Carrasco}
\affiliation{Departamento de F\'isica Te\'orica II, Universidad Complutense de Madrid, 28040 Madrid, Spain}
\author{J.~Nebreda}
\affiliation{Departamento de F\'isica Te\'orica II, Universidad Complutense de Madrid, 28040 Madrid, Spain}
\affiliation{Yukawa Institute for Theoretical Physics, Kyoto University, 606-8502 Kyoto, Japan} \author{J.R.~Pelaez}
\affiliation{Departamento de F\'isica Te\'orica II, Universidad Complutense de Madrid, 28040 Madrid, Spain}
\author{A.P.~Szczepaniak} 
\affiliation{Physics Department  Indiana University, Bloomington, IN 47405, USA }
\affiliation{Center for Exploration of Energy and Matter, Indiana University, Bloomington, IN 47403, USA}
\affiliation{Thomas Jefferson National Accelerator Facility, 
Newport News, VA 23606, USA}

\date{\today}

\begin{abstract}
We apply a recently developed dispersive formalism to calculate the Regge trajectories of the  
 $f_2(1270)$ and $f_2'(1525)$ mesons. Trajectories 
are calculated, not fitted to a family of resonances.
 Assuming that these spin-2 resonances can be treated in the elastic approximation the only input  
are the pole position and residue of the resonances. 
In both cases, the predicted Regge trajectories are almost real and linear, with slopes
in agreement with the universal value of order 1 GeV$^{-2}$. 
\end{abstract}
\maketitle

\section{Introduction}

There is  growing evidence for the existence of non-ordinary hadrons that do not follow
the quark model, {\it i.e.} the quark-antiquark-meson or three-quark-baryon classification.
Meson Regge trajectories relate resonance spins $J$ to the square of their masses  and for ordinary mesons  they are approximately linear.  The functional form of a  Regge trajectory depends on the 
  underlying dynamics and, for example,  the  linear trajectory for mesons is consistent 
  with the quark model as it can be explained in terms of a rotating relativistic flux tube that connects  the quark with the antiquark. 
  Regge trajectories associated with non-ordinary mesons do not, however, have to be linear. 
 The non-ordinary nature of the lightest scalar meson, the $f_0(500)$ also referred to as the $\sigma$, together with a few other scalars, has been postulated long ago \cite{Jaffe:1976ig}. 
  In the context of the  Regge classification, in a recent study of the meson spectrum  in 
\cite{Anisovich:2000kxa} it was concluded that the $\sigma$ meson does not belong to the same 
  set  of trajectories that many ordinary mesons do.
In \cite{Masjuan:2012gc}, it was concluded that the $\sigma$ can be omitted from the fits to  linear  $(J,M^2)$ trajectories  because of its large width. The reason is that its width was taken as measure of the uncertainty on its mass and it was found that, when fitting trajectory parameters,  its contribution to the overall $\chi^2$ 
was insignificant.

In a recent work \cite{Londergan:2013dza}  
we developed a  formalism based on dispersion relations that, instead of fitting  a specific, {\it e.g.} linear, form to spins and masses of various resonances, 
 enables us to calculate the trajectory using as input 
  the position and the residue of a complex resonance pole in a scattering amplitude. 
   When the method was applied  to the $\rho(770)$ resonance,
which appears as a pole in the elastic $P$-wave $\pi\pi$ scattering, the resulting trajectory was found to be, to a good approximation, linear. The resulting slope and intercept are in a  good agreement with phenomenological Regge fits.  The slope, which is slightly less than 1 GeV$^{-2}$, is expected to be universal for all ordinary trajectories.   It is worth noting that in this approach the resonance width is, as it should be, related to the imaginary part of the trajectory and not a source of an uncertainty. 
The $\sigma$ meson also appears as a pole in the $\pi\pi$ $S$-wave scattering. The position 
 and residue of the pole has recently been accurately determined in \cite{Caprini:2005zr} using rigorous dispersive formalisms.  \color{black}
 When the same method was applied to the
  $\sigma$ meson,  however,  we found quite a different trajectory. It has a significantly larger imaginary part and the slope parameter, computed at the physical mass as a derivative of the spin with respect to the mass squared, is more than one order of magnitude smaller than 
  the universal slope. The trajectory is far from linear, instead it is qualitatively similar to a trajectory
of  a Yukawa potential.   We also note that deviation  from linearity is not
   necessarily implied by the large width of the $\sigma$ since 
it was also shown in \cite{Londergan:2013dza} that resonances with large widths may 
 belong to linear trajectories. 
Our findings give further support  for the non-ordinary nature of the $\sigma$. 

Still, one may wonder if the single case of the  $\rho$ meson,  where the method agrees with 
 Regge phenomenology, gives sufficient evidence that it can distinguish between ordinary and non-ordinary mesons.  In this letter, therefore we show that other ordinary trajectories can be predicted with the same technique, as long as the underlying resonances are almost elastic. For this purpose,
we have concentrated on resonances that decay nearly $100\%$ to 
  two mesons. In addition to the $\rho$ there are two other well-known examples: the $f_2(1270)$, whose branching ratio to $\pi\pi$ is $84.8^{+2.4}_{-1.2}\%$, and the $f_2'(1525)$, with branching ratio to $K\bar K$ of $(88.7\pm2.2)\%$. These  resonances are well established in the quark model 
   and as we show below,   Regge trajectories predicted by our  method 
come out  almost real and linear with a slope close to the universal one. There is an additional check on 
 the method that we perform here.  Since the formalism used in the case of the $\rho$  
 was based on a twice-subtracted dispersion relation, the  trajectory had a linear term  plus a dispersive integral over the imaginary part. Since the imaginary part of the trajectory is closely related to the decay width, one might wonder if the $\rho(770)$, $f_2(1270)$
and $f_2'(1525)$  trajectories come out straight just because their widths are small.  
 In other words,  that for narrow resonances, the straight line behavior is not predicted but it is 
  already built in through subtractions.  For this reason, in this work, we also consider three subtractions and   show that for the ordinary resonances under study the quadratic term is negligible.

The paper is organized as follows. In the next section we briefly review the dispersive method and in 
 Sect.\ref{sec:numerical results} we  present the numerical results. 
  In Sect.\ref{sec:3subs} we  discuss results of the calculation with three subtractions.  
 Summary and outlook are given in Sect.\ref{conclusions}.

\section{Dispersive determination of a Regge trajectory from a single pole}

The partial wave expansion of the elastic scattering amplitude, $T(s,t)$, of two spinless 
 mesons of mass $m$ is given by 

\be
 T(s,t)=32 K \pi \sum_l (2l+1) t_l(s) P_l(z_s(t)),
\label{fullamp}
\ee
where $z_s(t)$ is the s-channel scattering angle and $K=1,2$ depending on whether the two 
 mesons are distinguishable or not. The  partial waves $t_l(s)$ are  normalized according to 
\be
t_l(s) =  e^{i\delta_l(s)}\sin{\delta_l(s)}/\rho(s), \quad \rho(s) = \sqrt{1-4m^2/s},
\ee
where $\delta_l(s)$ is the phase shift. The unitarity condition on the real axis in the elastic region, 
\be
\mbox{Im}t_l(s)=\rho(s)|t_l(s)|^2,
\label{pwunit}
\ee
is automatically satisfied.  When $t_l(s)$ is continued from the real axis to the entire complex plane, 
 unitarity determines the amplitude discontinuity across the cut on the real axis above $s=4m^2$. It 
   also determines the continuation in $s$, at fixed $l$, onto the second sheet  where resonance poles are located. 
  It follows from Regge theory that the same resonance poles appear when the amplitude is continued into the complex $l$-plane \cite{Reggeintro}, leading to 
  \be
t_l(s)  = \frac{\,\beta(s)}{l-\alpha(s)\,} + f(l,s),
\label{Reggeliket}
\ee
where $f(l,s)$ is analytical near $l=\alpha(s)$. The Regge trajectory $\alpha(s)$ and 
residue $\beta(s)$ satisfy  $\alpha(s^*)=\alpha^*(s)$, $\beta(s^*)=\beta^*(s)$, in the complex-$s$ plane cut along the real axis for $s > 4m^2$. 
Thus, as long as the pole dominates in Eq.\eqref{Reggeliket},
partial wave unitarity, Eq.\eqref{pwunit}, analytically continued to complex $l$ implies, 
\be
\mbox{Im}\,\alpha(s)   = \rho(s) \beta(s),   \label{unit} 
\ee
and determines the analytic continuation of  $\alpha(s)$  to the 
complex plane \cite{Chu:1969ga}. 
At threshold, partial waves behave as $t_l(s) \propto q^{2l}$, 
where $q^2=s/4-m^2$, so that if the Regge pole dominates the amplitude, we must have
$\beta(s) \propto q^{2\alpha(s)}$. 
Moreover, following Eq.\eqref{fullamp}, the Regge pole contribution to the full amplitude
is proportional to $(2\alpha + 1) P_\alpha(z_s)$, so that in order to cancel poles of the Legendre function $P_\alpha(z_s)\propto\Gamma(\alpha + 1/2)$ the residue has to vanish when 
  $\alpha + 3/2$ is a negative integer, {\it i.e.},
\be
\beta(s) =  \gamma(s) \hat s^{\alpha(s)} /\Gamma(\alpha(s) + 3/2).  \label{reduced} 
\ee
Here we defined $\hat s =( s-4m^2)/s_0$ and introduced a scale $s_0$ to have the right dimensions. 
 The so-called reduced residue, $\gamma(s)$, is a real 
analytic function. Hence, on the real axis above threshold, since $\beta(s)$ is real, the phase of $\gamma$ is
\be
\mbox{arg}\,\gamma(s) = - \mbox{Im}\alpha(s) \log(\hat s) + \arg \Gamma(\alpha(s) + 3/2). 
\ee
Consequently, we can write for $\gamma(s)$ a dispersion relation:
\be
\gamma(s) = P(s) \exp\left(c_0 + c' s + \frac{s}{\pi} \int_{4m^2}^\infty \!\!\!\!ds' \frac{\mbox{arg}\,\gamma(s')}{s' (s' - s)} \right), \label{g}
\ee
where $P(s)$ is an entire function. 
Note that the behavior at large $s$ cannot be determined from first principles, but, as we expect linear Regge trajectories for ordinary mesons, we should 
allow $\alpha$ to behave as a first order polynomial at large-$s$. 
This implies that $\mbox{Im}\alpha(s)$ decreases with growing $s$ 
and thus it obeys the dispersion relation~\cite{Reggeintro,Collins-PLB}: 
\be
\alpha(s) = \alpha_0 + \alpha' s + \frac{s}{\pi} \int_{4m^2}^\infty ds' \frac{ \mbox{Im}\alpha(s')}{s' (s' -s)}. \label{alphadisp}
\ee
Assuming $\alpha' \ne 0$, from  unitarity, Eq.\eqref{unit}, in order
to match the asymptotic behavior of $\beta(s)$ and $\mbox{Im}\alpha(s)$ it is required that  
$c' = \alpha' ( \log(\alpha'  s_0) - 1)$ and 
that $P(s)$ can at most be a constant, $P(s) = \mbox{const}$. 
Therefore, using Eq.\eqref{Reggeliket}, we arrive at the following three equations,
which define the ``constrained Regge-pole'' amplitude~\cite{Chu:1969ga}:
\begin{align}
\mbox{Re} \,\alpha(s) & =   \alpha_0 + \alpha' s +  \frac{s}{\pi} PV \int_{4m^2}^\infty ds' \frac{ \mbox{Im}\alpha(s')}{s' (s' -s)}, \label{iteration1}\\
\mbox{Im}\,\alpha(s)&=  \frac{ \rho(s)  b_0 \hat s^{\alpha_0 + \alpha' s} }{|\Gamma(\alpha(s) + \frac{3}{2})|}
 \exp\Bigg( - \alpha' s[1-\log(\alpha' s_0)] 
+  \!\frac{s}{\pi} PV\!\!\!\int_{4m^2}^\infty\!\!\!\!\!\!\!ds' \frac{ \mbox{Im}\alpha(s') \log\frac{\hat s}{\hat s'} + \mbox{arg }\Gamma\left(\alpha(s')+\frac{3}{2}\right)}{s' (s' - s)} \Bigg), 
\label{iteration2}\\
 \beta(s) &=    \frac{ b_0\hat s^{\alpha_0 + \alpha' s}}{\Gamma(\alpha(s) + \frac{3}{2})} 
 \exp\Bigg( -\alpha' s[1-\log(\alpha' s_0)] 
+  \frac{s}{\pi} \int_{4m^2}^\infty \!\!\!\!\!\!\!ds' \frac{  \mbox{Im}\alpha(s') \log\frac{\hat s}{\hat s'}  + \mbox{arg }\Gamma\left(\alpha(s')+\frac{3}{2}\right)}{s' (s' - s)} \Bigg),
 \label{betafromalpha}
 \end{align}
where $PV$ denotes the principal value.  For real $s$, the last two equations reduce to Eq.\eqref{unit}.
The three equations are solved numerically with the free parameters fixed by demanding 
 that the pole on the second sheet of the amplitude in Eq.~(\ref{Reggeliket}) is at a given location. 
Thus we will be able to obtain the two independent trajectories corresponding to the  $f_2(1270)$ and $f_2'(1525)$ resonances from their respective pole parameters. Note that we are not imposing, but just allowing, linear trajectories.

\section{Numerical Results}
\label{sec:numerical results}
 
In principle,  the method described in the previous section
is suitable for resonances that appear in
the  elastic scattering amplitude, {\it i.e.} they only decay to one two-body channel.
For simplicity we are also focusing on cases were the two mesons 
in the scattering state have the same mass. 
We assume that both the $f_2(1270)$  and the $f_2'(1525)$ resonances 
 can be treated as purely elastic and we will use their decay fractions into channels other than $\pi\pi$ and $K\bar K$, respectively, as an additional systematic uncertainty in their widths and couplings. 
In our numerical analysis we fit the pole, $s_p$,  and residue, $|g^2|$,  
found in the second Riemann sheet of the Regge amplitude.
In this amplitude the
$\alpha(s)$ and $\beta(s)$ are constrained to satisfy
the dispersion relations in Eqs.\eqref{iteration2} and  \eqref{betafromalpha}.
Thus, the fit determines the parameters  $\alpha_0, \alpha',b_0$ 
  for  the trajectory of each resonance. 
In practice, we minimize  the sum of squared differences between the input and output values 
for the real and imaginary parts of the pole position and for the absolute value 
of the squared coupling, divided by the square of the corresponding uncertainties. 
At each step in the minimization procedure a set of $\alpha_0, \alpha'$ and $b_0$ parameters is chosen and the system of Eqs.\eqref{iteration1} 
and \eqref{iteration2} is solved iteratively. The resulting Regge amplitude for each $\alpha_0, \alpha'$ 
and $b_0$ is then continued to the 
complex plane, in order to determine the resonance pole in the second Riemann sheet, 
and  the $\chi^2$ is calculated by comparing this pole to the corresponding input.

\subsection{$f_2(1270)$ resonance}
\label{subsec:f2(1270)}

 In the case of the $f_2(1270)$ resonance, we use as input the pole obtained from the conformal parameterization of the D0 wave from Ref~\cite{GarciaMartin:2011cn}. In that work the authors use different parameterizations in different regions of energy and impose a matching condition. Here we will  use the parameterization valid in the region where the resonance dominates the scattering amplitude, namely, in the interval $2m_K\le s^{1/2}\le 1420$ MeV. Moreover, we will decrease the width down to 
  $85\%$ of the value found in \cite{GarciaMartin:2011cn} to account for the inelastic  
   channels. The conformal parameterization results in the pole located at 
 $$\sqrt{s_{f_2}}=M-i\Gamma/2=1267.3^{+0.8}_{-0.9}-i(87\pm9)\text{ MeV}$$ 
 and a coupling of 
  $$|g_{f_2\pi\pi}|^2=25\pm3\text{ GeV}^{-2}.$$

\begin{figure}
\hspace*{-.6cm}
\includegraphics[scale=0.9,angle=-90]{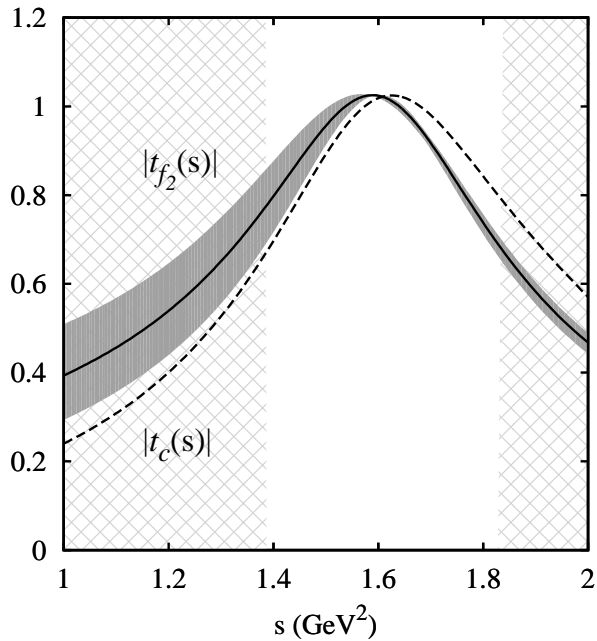}
 \caption{\rm \label{Fig:ampl_f2} 
The solid line represents the absolute value of the constrained Regge-pole amplitude for the $f_2(1270)$ resonance. The gray bands cover the uncertainties due to the errors in the input pole parameters. The dashed line corresponds to the absolute value of the data fit obtained in \cite{GarciaMartin:2011cn}. Let us recall that only the parameters of the pole given by this parameterization have been used as input, and not the amplitude itself.
The regions covered with a mesh correspond to $s<(M-\Gamma/2)^2$ and $s>(M+\Gamma/2)^2$,
where the background might not be negligible anymore. }
\end{figure}

With these input parameters, we follow the minimization procedure as explained above, until we get a Regge pole at  $\sqrt{s_{f_2}}=(1267.3\pm0.9)-i(89\pm10)\text{ MeV}$ and coupling $|g_{f_2\pi\pi}|^2=25\pm3\text{ GeV}^{-2}$.  In Fig.~\ref{Fig:ampl_f2} we show the corresponding constrained Regge-pole amplitude on the real axis versus the conformal parameterization that was  constrained by the data  \cite{GarciaMartin:2011cn}. 
This comparison is a check that our Regge-pole amplitude, which neglects the background $f(l,s)$ term 
in Eq.\eqref{Reggeliket}, describes well the amplitude in the pole region,
namely for $(M-\Gamma/2)^2<s<(M+\Gamma/2)^2$.
The grey bands cover the uncertainties arising from the errors of the input and include an additional 
 $15\%$ systematic  uncertainty in the width as explained above. 
Taking into account that
only parameters of the pole have been fitted, but not the whole amplitude in the real axis, and that we have completely neglected the background in Eq.~(\ref{Reggeliket}), the agreement between  the two amplitude models is very good, particularly in the resonance region.  Of course, the agreement deteriorates as we move away from the peak region as illustrated  by the shadowed energy regions $s<(M-\Gamma/2)^2$ and $s>(M+\Gamma/2)^2$. 

 Since our constrained Regge amplitude provides a good description of the resonance region we can trust the resulting Regge trajectory. The parameters of the trajectory obtained through our minimization procedure are as follows, 
 \begin{equation}
\alpha_0=0.9^{+0.2}_{-0.3}\,;\hspace{3mm} \alpha'=0.7^{+0.3}_{-0.2} \text{ GeV}^{-2};\hspace{3mm}b_0=1.3^{+1.4}_{-0.8}\,.\label{eq:paramf2}
\end{equation}

\begin{figure}
\hspace*{-.6cm}
\includegraphics[scale=0.9,angle=-90]{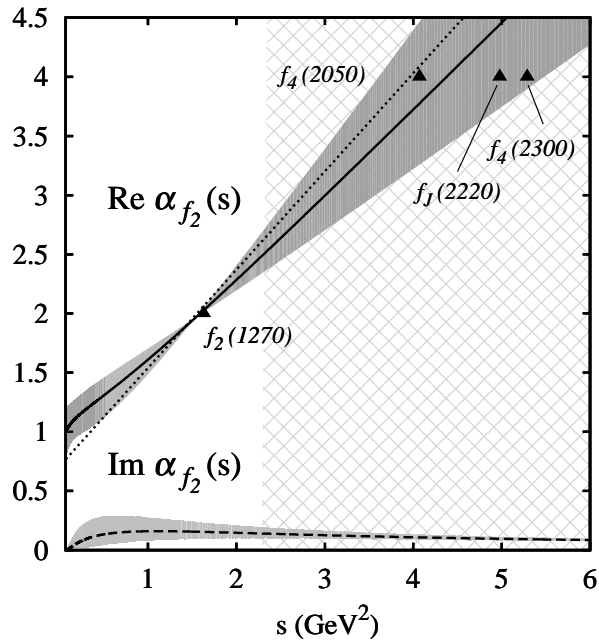}
 \caption{\rm \label{Fig:alpha_f2} 
Real (solid) and imaginary (dashed) parts of the $f_2(1270)$ Regge trajectory.
 The gray bands cover the uncertainties due to the errors in the input pole parameters.  The area 
covered with a mesh is the mass region starting
three half-widths above the resonance mass, where our elastic approach should be considered only as a mere extrapolation. For comparison, we show with a dotted line the $f_2(1270)$ Regge trajectory obtained in~\cite{Anisovich:2000kxa}, traditionally called the $P'$ trajectory. We also show the resonances listed in the PDG that are candidates for this trajectory. Note that their average mass does not always coincide with the nominal one, as is the case for the $f_2(1270)$.}
\end{figure}

In Fig.~\ref{Fig:alpha_f2} we show the real and imaginary parts of $\alpha(s)$, with solid and dashed lines, respectively. Again, the gray bands cover the uncertainties coming from the errors in the input pole parameters. We find that the real part of the trajectory is almost linear and much bigger than the imaginary part. It is as expected for Regge trajectories of ordinary mesons.  For comparison, we also show, with a dotted line, the Regge trajectory obtained in~\cite{Anisovich:2000kxa} by fitting a Regge linear trajectory to the 
meson states associated with $f_2(1270)$, which is traditionally referred to as the 
$P'$ trajectory. We see that the two trajectories are in good agreement. Indeed, our parameters are compatible, within errors, with those in~\cite{Anisovich:2000kxa}: $\alpha_{P'}\approx0.71$ and $\alpha'_{P'}\approx0.83\text{ GeV}^{-2}$. 
We also include in Fig.~\ref{Fig:alpha_f2} the resonances from the PDG~\cite{PDG} listing that could be associated with this trajectory.

In Fig.~\ref{Fig:alpha_f2} the trajectory has been extrapolated to  high energies, 
where the elastic approximation does not hold any more and we cannot hope to give a precise prediction for its behavior. 
The only reason to do this is to show the position of the candidate states 
 connected to  the $f_2(1270)$. 
In the figure, this region is covered with a mesh to the right 
 from the line at the $s$ that corresponds to the resonance mass plus 
 three half-widths. 
Of course, we cannot confirm
which of these resonances belongs to the $f_2(1270)$ trajectory, 
but we observe that the $J=4$ resonance could be the $f_4(2050)$, as proposed in~\cite{Anisovich:2000kxa}, or the $f_J(2220)$
\footnote{This resonance still ``needs confirmation'' and it is not yet known whether its spin is 2 or 4 \cite{PDG}.} or even the $f_4(2300)$. All these resonances appear in the PDG, but are  omitted from the summary tables. 

\subsection{$f_2'(1525)$ resonance}
\label{subsec:f2'(1525)}
  
As commented above, the $f_2'(1525)$ decays mainly to two kaons. Although there is 
no scattering data on the  $l=2$ elastic $\bar KK$ phase shift in this mass region, 
the mass and width of the $f_2'(1525)$ are given in the PDG~\cite{PDG}. Thus we use
$M_{f_2'}=1525\pm5$ MeV and $\Gamma^{KK}_{f_2'}=69^{+10}_{-9}$ MeV, 
where the central value of this width corresponds to the 
decay into $\bar KK$ only. 
Now, we infer the scattering pole parameters assuming the $f_2'(1525)$
is well described by an elastic Breit-Wigner shape, so that we take the pole
to be at $s_{f_2'}=(M_{f_2'}-i \Gamma_{f_2'}/2)^2$ and the
residue to be ${\rm Res}=-M_{f_2'}^2\Gamma_{f_2'}^{KK}/2p$, where $p$ is the CM momenta of the two kaons.
Since $\vert g \vert^2=-16 \pi (2l+1)\,{\rm Res}/(2p)^{2l}$, we find $|g_{f_2'KK}|^2=19\pm 3 \text{ GeV}^{-2}$.

\begin{figure}
\hspace*{-.6cm}
\includegraphics[scale=0.9,angle=-90]{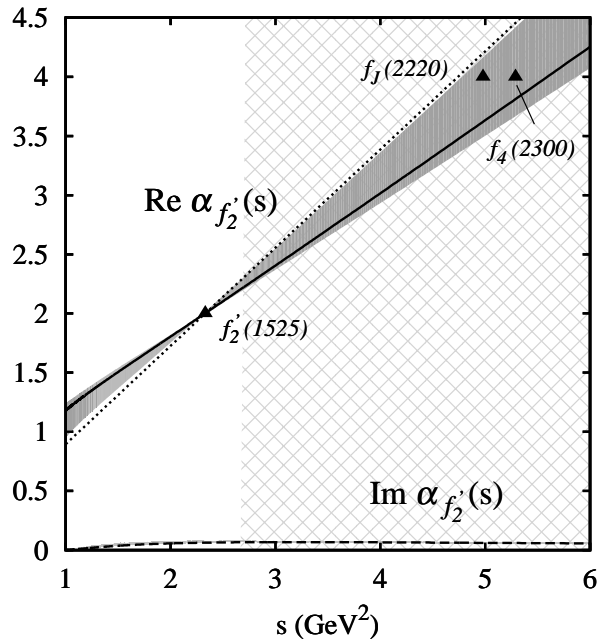}
 \caption{\rm \label{Fig:alpha_ff2} 
Real (solid) and imaginary (dashed) parts of the $f_2'(1525)$ Regge trajectory.
 The gray bands cover the uncertainties due to the errors in the input pole parameters and 
the  area covered with a mesh is the mass region starting three half-widths above the resonance mass, where our elastic approach must be considered just as an extrapolation. For comparison, we show with a dotted line the Regge trajectory obtained in~\cite{Anisovich:2000kxa} and the resonances listed in the PDG that could belong to this trajectory.}
\end{figure}

With these input parameters we solve the dispersion relations using 
the same minimization method and obtain the following Regge pole parameters: $\sqrt{s_{f_2'}}=(1525\pm5
)-i(34^{+4}_{-5})\text{ MeV}$ and $|g_{f_2'KK}|^2=19\pm3\text{ GeV}^{-2}$. Since we lack 
  experimental data to compare the amplitudes, we proceed to  examining the trajectory. The parameters that we obtain are, 
\begin{equation}
\alpha_0=0.53^{+0.10}_{-0.44}\,;\hspace{3mm} \alpha'=0.63^{+0.20}_{-0.05} \text{ GeV}^{-2};\hspace{3mm}b_0=1.33^{+0.63}_{-0.09}\,,\label{eq:paramsf2p}
\end{equation}
which give the Regge trajectory shown in Fig.~\ref{Fig:alpha_ff2}. Again, we find the real part nearly linear and much larger than the imaginary part. As in the case of the $f_2(1270)$, the slope is compatible with that found for the $P'$ trajectory in~\cite{Anisovich:2000kxa} $\alpha'_{P'}\approx0.83\text{ GeV}^{-2}$, 
and the intercepts also agree. 
  
  As we did for $f_2(1270)$, we include in Fig.~\ref{Fig:alpha_ff2} the $J=4$ candidates for the $f_2'(1525)$ trajectory. These are the $f_J(2220)$ and the $f_4(2300)$. 
   We remark that  there is no experimental evidence of the $f_4(2150)$ that was predicted in~\cite{Anisovich:2000kxa} from their analysis of the $f_2'(1525)$ trajectory. As commented before, these resonances 
lie in a region, covered with a mesh in Fig.~\ref{Fig:alpha_ff2},
 beyond the strict applicability limit of our approach, 
where our results must  be considered qualitatively at most.
  
  Finally, we remark that the PDG list  includes another $f_2$ resonance, albeit requiring 
   confirmation. It has a mass between that of the $f_2(1270)$ and the $f_2'(1525)$ and 
     it could also have  either the $f_J(2220)$ or $f_4(2300)$ as the higher mass partner.

\section{Dispersion relation with three subtractions} 
\label{sec:3subs}

As already mentioned in the introduction, one may wonder whether the
 linearity of the trajectories we obtain for the two D-wave resonances,   as well as for the $\rho(770)$ in~\cite{Londergan:2013dza}, is related to use of two subtractions 
  in the dispersion relation  for $\alpha(s)$. In particular,  since the resonances are rather narrow 
one could  expect the imaginary part of 
their trajectories to be small, so that if the last term in Eq.~\eqref{alphadisp} was
dropped the trajectory would be reduced to a straight line. Thus, in order to show that the linearity 
of the trajectory is not forced by the particular parameterization, 
we repeated the calculations using three subtractions in the dispersion relations, 

\begin{align}
\mbox{Re} \,\alpha(s) & =   \alpha_0 + \alpha' s + \alpha'' s^2  + \frac{s^2}{\pi} PV \int_{4m^2}^\infty ds' \frac{ \mbox{Im}\alpha(s')}{s'^2 (s' -s)}, \label{iteration1-3sub}\\
\mbox{Im}\,\alpha(s)&=  \frac{ \rho(s)  b_0 \hat s^{\alpha_0 + \alpha' s+\alpha'' s^2} }{|\sqrt{\Gamma(\alpha(s) + \frac{3}{2})}|}
 \exp\Bigg( -  \frac{1}{2}[1-\log(\alpha'' s_0^2)] s (R+\alpha'' s)-Q s\nonumber\\ 
 &  \hspace{4.4cm}
+  \!\frac{s^2}{\pi} PV\!\!\!\int_{4m^2}^\infty ds' \frac{ \mbox{Im}\alpha(s') \log\frac{\hat s}{\hat s'} + \frac{1}{2}\mbox{arg }\Gamma\left(\alpha(s')+\frac{3}{2}\right)}{s'^2 (s' - s)} \Bigg), 
\label{iteration2-3sub}
 \end{align}
 with
 \begin{equation}
 R=B-\frac1{\pi}\int_{4m^2}^\infty ds' \frac{ \mbox{Im}\alpha(s')}{s'^2},
 \end{equation}
 and
 \begin{equation}
 Q=-\frac1{\pi}\int_{4m^2}^\infty ds' \frac{ -\mbox{Im}\alpha(s')\log{\hat s'}+\frac{1}{2}\mbox{arg }\Gamma\left(\alpha(s')+\frac{3}{2}\right)}{s'^2}.
 \end{equation}
 
The reason why the constants $R$ and $Q$ and the square root of $\Gamma$ have been introduced  is to ensure that, at large $s$, $\mbox{Im}\,\alpha(s)$ behaves as $1/s$. The parameters that we obtain for the trajectories with these dispersion relations  are shown in Table~\ref{Tab:3sub}.


\renewcommand{\arraystretch}{1.3}
\begin{table}
\centering
\caption{Parameters of the $f_2(1270)$, $f_2'(1525)$  and $\rho(770)$ Regge trajectories using three-time-subtracted dispersion relations.} \label{Tab:3sub}
\begin{tabular*}{0.7\textwidth}{@{\extracolsep{\fill} }ccccc}\hline
& $\alpha_0$ & $\alpha'$ (GeV$^{-2}$) &  $\alpha''$ (GeV$^{-4}$)  & \hspace{5mm}$b_0$\hspace{5mm} \\\hline
$f_2(1270)$ &  1.01 &  0.97 &   0.04  & 2.13\\
$f_2'(1525)$ &   0.42   &    0.65       &  0.02  & 4.58 \\
$\rho(770)$ &   0.56 &   1.11  &   0.03  &  0.88 \\
\hline
\end{tabular*}
\end{table}

\begin{figure}
\begin{tabular}{ccc}
\includegraphics[scale=0.6,angle=-90]{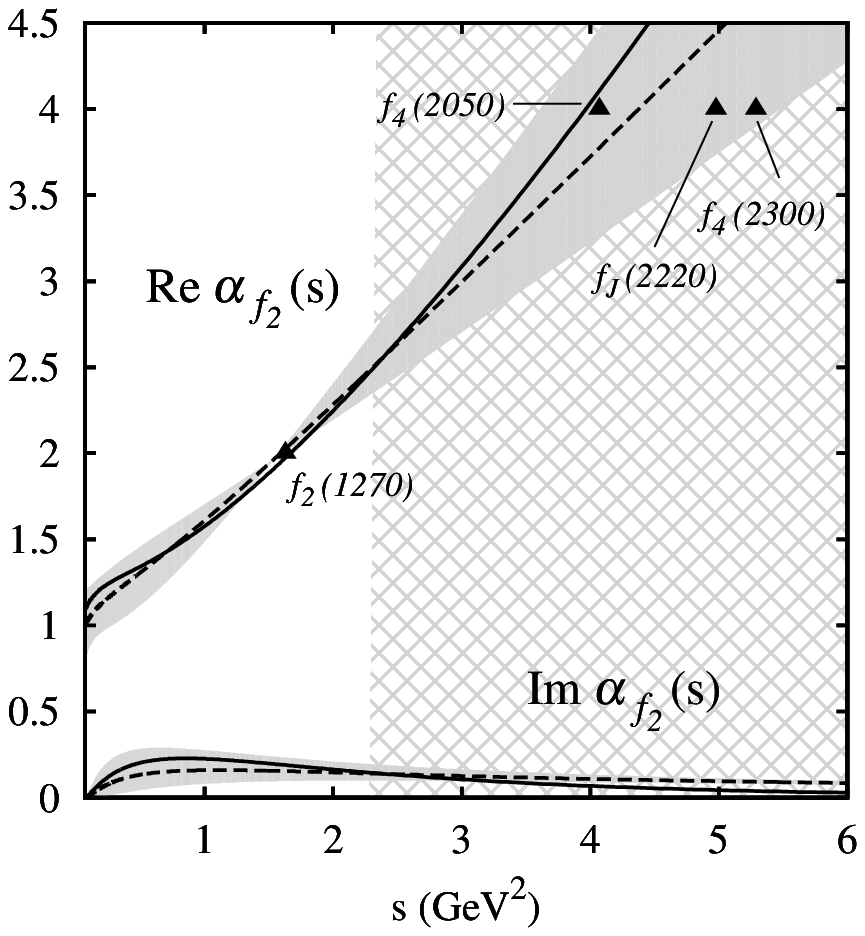}&
\hspace{-3mm}\includegraphics[scale=0.6,angle=-90]{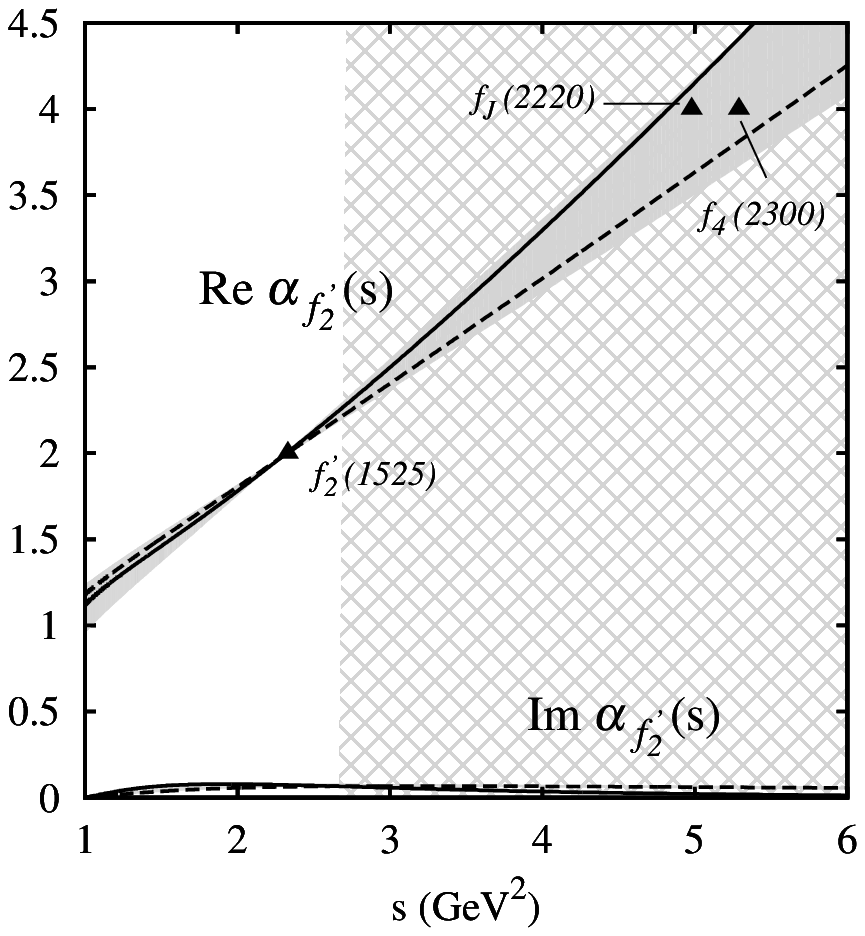}&
\hspace{-3mm}\includegraphics[scale=0.6,angle=-90]{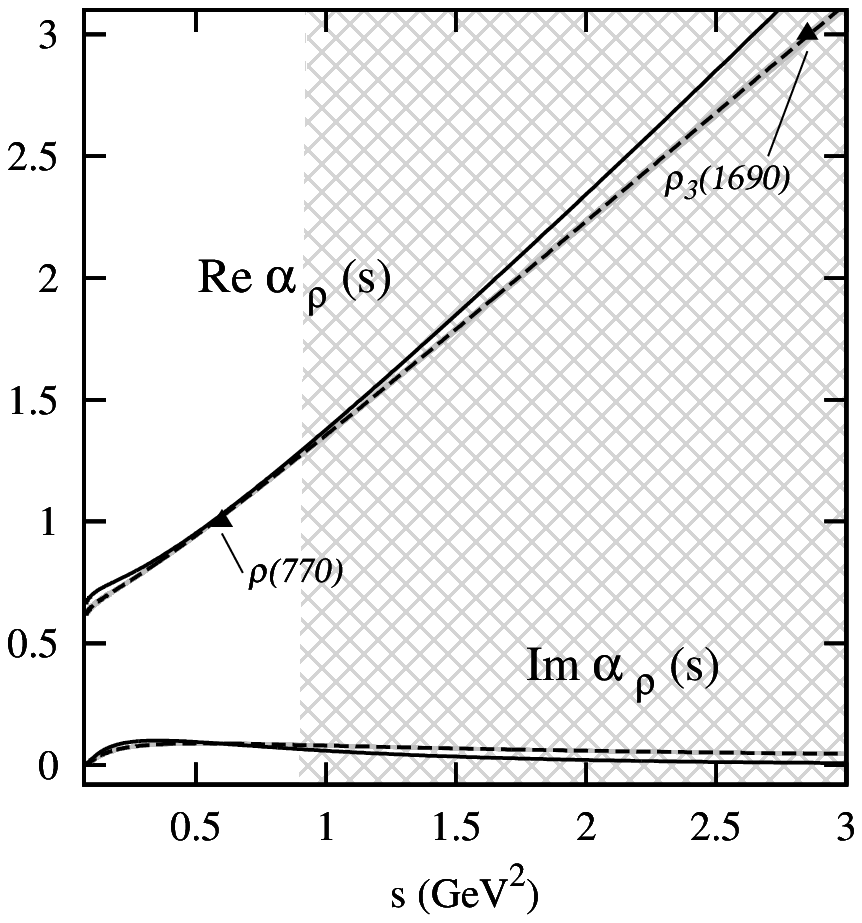}
\end{tabular}
 \caption{\rm \label{Fig:3sub} Regge trajectories obtained using three-time-subtracted dispersion relations (solid lines) compared to the ones obtained with twice-subtracted dispersion relations (dashed lines with gray error bands).}
\end{figure}

With the above parameterization we obtain for the fitted pole parameters 
 $\sqrt{s_{f_2}}=1267.3-i 90 \text{ MeV}$,  $|g_{f_2\pi\pi}|^2=25\text{ GeV}^{-2}$, $\sqrt{s_{f_2'}}=1525-i35\text{ MeV}$,   $|g_{f_2'\pi\pi}|^2=19\text{ GeV}^{-2}$,  $\sqrt{s_{\rho}}=763-i 74\text{ MeV}$ and $|g_{\rho\pi\pi}|^2=35\text{ GeV}^{-2}$.  Therefore, 
despite having four parameters to fit three numbers, we find no real improvement in the description of the poles. 
In the case of three subtractions, neglecting the imaginary part of the resonances 
 results in a quadratic trajectory.   
 Therefore, in Fig.~\ref{Fig:3sub} we compare the trajectories using the  three (solid line) and the two 
   (dashed line) subtractions in the  dispersion relations. We observe that in both cases 
    these is a  curvature, but that in the elastic region the trajectories are  almost linear. 
    The difference between the two methods only becomes apparent for masses well above the range of applicability.  Moreover, the difference between the results obtained using 
 two and three subtractions can be used as an indicator of the stability of our results and therefore confirms that the applicability range for our method is well estimated and ends as soon as the inelasticity in the wave becomes sizable.

\section{Discussion, conclusions and outlook}
\label{conclusions}
In~\cite{Londergan:2013dza} a dispersive method was developed to 
calculate Regge trajectories of resonances that appear in the elastic scattering of
two mesons. We showed how, using the associated scattering pole of the resonance
it is possible to determine whether its trajectory is of a  standard type, {\it i.e.} real and linear
as followed by ``ordinary''  $\bar qq$-mesons, or not.
This method thus provides a possible benchmark for identifying non-ordinary mesons.
In particular the ordinary Regge trajectory of the $\rho(770)$, which is a well-established $\bar qq$ state, was successfully predicted, whereas the $\sigma$ meson, 
a long-term candidate for a non-ordinary meson, was found to follow a completely different trajectory.


In the first part of this work we have successfully predicted the trajectories of other two,  well-established ordinary resonances,  the $f_2(1270)$ and $f_2'(1525)$. In particular, from parameters of the associated poles in the  complex energy plane we have calculated their trajectories and 
 have shown that they are almost real and very close to a straight line, as expected.  
 
In the second part of this work we have addressed the question of whether
choosing two subtractions in the dispersion relations of \cite{Londergan:2013dza}
was actually imposing that the real part of the trajectory is a straight line 
for relatively narrow resonances. 
To address this question we analyzed the same resonances using a dispersion relation with an   additional subtraction.  We have shown that within the range  of applicability of our approach, which basically coincides with the elastic regime, the resulting trajectories are once again very close to a straight line. 

In the future it will be interesting to use such dispersive methods to determine trajectories of other mesons 
{\it e.g} the   $K^*(892)$  as well as the controversial ``partner" the scalar $K^*(800)$, which 
is another long-time candidate for a non-ordinary meson. Heavy mesons in charm and beauty sectors can also be examined. 
 We also plan to extend the method  to meson-baryon scattering, where, for example, the $\Delta(1232)$ is another candidate for an 
  ordinary resonance. We are also extending the approach to coupled channels. 

{\bf Acknowledgments} 
We would like to thank M.R. Pennington for several discussions. 
JRP and JN are supported by the Spanish project FPA2011-27853-C02-02. JN acknowledges funding by the Fundaci\'on Ram\'on Areces. 
APS work is supported in part by the U.S. Department of Energy, Office of Science, Office of Nuclear Physics under contract DE-AC05-06OR23177 and DE-FG0287ER40365.

APS\ is supported in part by the U.S.\ Department of Energy under Grant DE-FG0287ER40365.

\end{document}